\documentclass[12pt]{article}
\usepackage[T2A]{fontenc}
\usepackage[cp1251]{inputenc}
\usepackage{hyperref}
\usepackage{a4wide}
\usepackage{amsmath,latexsym,amsthm,amsfonts}
\usepackage{amssymb,amscd}
\usepackage{graphicx}
\def\cnt#1#2{\text{\setbox2=\hbox{#1}\rlap{\hbox to
\wd2{\hfil#2\hfil}}\box2}}
\def\SI{\mathbin{\cnt{$\sum$}{$\int$}}}

\newcommand{\R}{\Bbb R}

\newcommand{\X}{\frak X}
\newcommand{\bml}[1]{\begin{multline}\label{#1}}
\newcommand{\bee}{\begin{equation}}
\newcommand{\bed}{\begin{displaymath}}
\newcommand{\ee}{\end{equation}}

\newcommand{\bs}{\begin{split}}

\newcommand{\be}{\beta}
\newcommand{\ga}{\gamma} \newcommand{\Ga}{\Gamma}
 \newcommand{\De}{\Delta}
\newcommand{\la}{\lambda} \newcommand{\La}{\Lambda}

\newcommand{\si}{\sigma}

 \newcommand{\sm}{\setminus}

\newcommand{\w}{\widetilde}

\newtheorem{definition}{Definition}
\newtheorem{theorem}{Theorem}[section]
\newtheorem{lem}{Lemma}[section]

\newtheorem{prop}{Proposition}[section]
\newtheorem{remark}{Remark}[section]
\newtheorem{corollary}{Corollary}[section]
\theoremstyle{definition} 

\theoremstyle{remark} 
 \numberwithin{equation}{section}
\begin{document}

\title{Quasi-lattice approximation of statistical systems with strong superstable interactions. Correlation functions}
\author{A.~L.~Rebenko$^1$, M.~V.~Tertychnyi $^2$}
\date{}
\maketitle
\begin{footnotesize}
\begin{tabbing}
$^1$ \= Institute of Mathematics, Ukrainian National Academy of Sciences, Kyiv, Ukraine\\
\> {\em rebenko@voliacable.com  ; rebenko@imath.kiev.ua}\\
$^2$ \> Faculty of mechanics and mathematics, Kyiv Shevchenko university, Kyiv, Ukraine \\
\>{\em mt4@ukr.net}\\
\end{tabbing}
\end{footnotesize}
\begin{abstract}
A continuous infinite system of point particles interacting via
two-body  strong superstable potential is considered in the
framework of classical statistical mechanics. We define some kind of
approximation of  main quantities, which describe
 macroscopical and microscopical characteristics of  systems,
 such as grand partition function   and correlation functions.
The pressure   of an approximated system converge to the pressure of
the initial system if the parameter of approximation $a\rightarrow
0$ for any values of an inverse temperature $\beta
>0$ and a chemical activity $z$. The same result is true for the
family of correlation functions in the region of small z.
\end{abstract}
\thispagestyle{empty}



\section{\bf Introduction}
The main achievements of mathematical physics in research of
critical phenomena are connected first of all with studying infinite
lattice systems. But one can see  totally another situation
concerning continuous systems. The mathematical results have been
obtained in the majority of cases only for the small values of
parameters $\beta = \frac{1}{kT}$ \,( where $T$ is a temperature)
and a chemical  activity $z$. The research of continuous systems in
the area of critical values of these parameters is restricted to
some artificial models like the Widom-Rowlinson model \cite{WR70} or
with field theory of type Hamiltonian \cite{LMP99}, and the methods
of investigation  are copied from lattice systems(see, {\it e.g.},
\cite{Ru71}, \cite{LL72}, using Peierls' argument, \cite{BKL84},
using Pirogov-Sinai theory or \cite{GH96},\cite{GHM00}, using random
cluster expansion). Another type of arguments was invented by Gruber
and Griffiths  \cite{GG86} and  used in \cite{RZ98},\cite{GTZ02} to
prove the existence of orientational ordering transitions
 in the continuous-spin models of ferrofluid.

 Some important characteristics
 of critical phenomena can be also
 described by using lattice approximation of  continuous
 systems. It was especially  successful to apply lattice
 approximation to research of the models of  quantum field
 theory (see, {\it e.g.}, \cite{Si74} and references therein).
Substantial progress was also reached in studying
 models of lattice-gas(\cite{Ru69}). But the main disadvantage of
 the last example is that it does not contain the parameter that ensures
 the transition to the classical continuous gas.

 On the other hand the main mathematical problems in the research
 of infinite  continuous systems appear because  it is necessary to take into account all
 possible configurations of particles, even if the probability of
 their occurrence is rather small. One of possible ways to solve this
 problem is to introduce hard-core potentials. It helps to avoid
  mathematical difficulties which is connected with an accumulation
  of many number of particles in the small volume , but at the same time it
  leads to some new problems, that is connected with interpretation of
 physical results and application of some mathematical methods.

 In the present article we propose some intermediate approximation
 of several main quantities, which describe
 macroscopical and microscopical characteristics of  systems,
 such as grand partition function   and correlation functions.
 The main idea is in the following:  we split the space $\mathbb{R}^d$ into
 nonintersecting hyper cubes with a volume $a^d$ and define approximated
 grand partition function and the family of approximated correlation
 functions in such a way, that they take into account only such
 configurations of particles in $\mathbb{R}^d$, when there is not
 more than one particle in each cube.

 It was shown in this work, that for the potentials which have
 non integrable singularity in the neighborhood of the origin(strong superstable
 potentials) the pressure   of the approximated
 system converge to the pressure  of the
 initial system if $a\rightarrow 0$ for any values of an inverse temperature $\beta >0$ and a chemical activity
 $z$. The same
result is true for the family of correlation functions in the region
of small z.

\section{\bf Notations and main results}
\subsection{\bf Configuration space}
 Let ${{\Bbb R}}^{d}$ be a $d$-dimensional Euclidean
space. The set of positions $\{x_i\}_{i\in{\Bbb N}} $ of identical
particles is considered to be  a locally finite subset in ${\Bbb
R}^d$ and the set of all such subsets creates the configuration
space:
\begin{equation*}
\Gamma=\Gamma_{{\Bbb R}^{d}} :=\left\{ \left. \gamma \subset {{\Bbb
R}}^{d}\right| \,|\gamma \cap \Lambda
|<\infty,\,\,\mathrm{for}\,\,\mathrm{all}\; \Lambda \in
\mathcal{B}_c({{\Bbb R}}^{d})\right\},
\end{equation*}
where $|A|$ denotes the cardinality of the set $A$ and
$\mathcal{B}_{c}({{\Bbb R}}^{d})$ denote the systems of all bounded
Borel sets in ${\Bbb R}^d$. We also need to define the space of
finite configurations $\Gamma_{0}$:
$$
\Gamma_{0} = \bigsqcup_{n\in {\Bbb N}_0}\Gamma^{(n)},\quad
\Gamma^{(n)}:= \{\eta\subset{\Bbb R}^{d}\; |\; |\eta|=n \},\quad
{\Bbb N}_0={\Bbb N}\cup \{0\}.
$$

 For every $\Lambda \in \mathcal{B}_c({{\Bbb R}}^{d} )$ one
can define a mapping $N_\Lambda :\Gamma\rightarrow \Bbb{N}_0$ of the
form
$$N_\Lambda (\gamma):=|\gamma \cap \Lambda |\;=\;|\ga_\La|.$$ The Borel $\sigma
$-algebra $ {\frak B}(\Gamma)$ is equal to $\sigma (N_\Lambda \left|
\Lambda \in {\frak B}_c({{\Bbb R}}^{d})\right.)$ and additionally
one may introduce the following filtration $${\frak B}_\Lambda
(\Gamma ):=\sigma (N_{\Lambda ^{\prime }}\left| \Lambda ^{\prime
}\in \mathcal{B}_c({{\Bbb R}}^{d}),\,\,\Lambda ^{\prime }\subset
\Lambda \right. ),$$ see \cite{Le75I}, \cite{Le75II}, \cite{AKR97}
for details.

We need also to define

$$
\Gamma_{\La} :=\left\{ \left. \eta \in \Ga_0\, \right| \,\eta
\subset \Lambda  \right\}.
$$

By ${\frak B}(\Gamma_\Lambda)$ we denote the corresponding
$\sigma$-algebra on $\Gamma_\Lambda$. For the given intensity
measure $\sigma$ (in this context $\sigma$ is Lebesgue measure on
$\mathcal{B}({\Bbb R}^d)$) and any $n\in \Bbb{N}$ the product
measure $\sigma ^{\otimes n}$ can be considered as a measure on
$$\widetilde{({{\Bbb R}}^{d})^n}=
\left\{ \left. (x_1,\ldots ,x_n)\in ({{\Bbb R}}^{d})^n\right| \,x_k\neq x_l\,\,\mathrm{if}%
\,\,k\neq l\right\}$$ and hence as a measure $\sigma ^{(n)}$ on
$\Gamma^{(n)}$ through the map
$$\:sym_n:\widetilde{({\Bbb R}^{d})^{n}}\ni
(x_1,...,x_n)\mapsto\{x_1,...,x_n\}\in\Gamma^{(n)}.$$

Define the Lebesgue-Poisson measure $\lambda_{z\sigma}$ on ${\frak
B}(\Gamma_{0})$ by the formula:
\begin{equation}\label{21}%(2.1)
 \lambda_{z\sigma} :=\sum_{n\ge 0}\frac{z^n}{n!}\sigma^{(n)}.
\end{equation}

The restriction of $\lambda_{z\sigma}$ to ${\frak
B}(\Gamma_\Lambda)$ we also denote by $\lambda_{z\sigma}$. For  more
detailed structure of the configuration spaces $\Gamma$, $\Gamma_0$,
$\Gamma_\Lambda$ see \cite{AKR97}.

As in \cite{Re98} define two additional configuration spaces: a
space of {\it dilute} configurations and  a space of {\it dense }
configurations.

Let $a>0$ be arbitrary. Following \cite{Ru70} for each $r\in{\Bbb
Z}^{d}$ we define an elementary cube with an edge $a$ and a center
$r$
\begin{equation}\label{22}%(2.2)
\Delta_{a}(r):=\{x\in{\Bbb R}^d\mid a(r^i-1/2)\leq x^i<a(r^i+1/2)\}.
\end{equation}
 We will  write $\Delta$ instead of $\Delta_{a}(r)$,
if a cube $\Delta$ is considered to be arbitrary and there is no
reason to emphasize that it is centered at the concrete point
$r\in{\Bbb Z}^{d}$. Let $\overline\Delta_a$ be the partition of
$\mathbb{R}^d$ into cubes $\Delta_{a}(r)$. Without loss of
generality  consider only that $\Lambda \in \mathcal{B}_c({{\Bbb
R}}^{d})$ which is union of cubes $\Delta_{a}(r)$. Then for any
$X\subseteqq\La$ which is a union of cubes $\De\in\overline\Delta_a$
define

\begin{equation}\label{23}%(2.3)
\Gamma^{dil}_{X} :=\left\{ \left. \gamma \in\Gamma_{X} \right|
\,N_\De(\ga)=0 \vee 1 \; \text{for all}\; \Delta\subset X \right\}
\end{equation}

and

\begin{equation}\label{24}%(2.4)
\Gamma^{den}_{X} :=\left\{ \left. \gamma \in\Gamma_{X} \right|
\,N_\De(\ga)\geq2 \; \text{for all}\; \Delta\subset X \right\}.
\end{equation}

\subsection{\bf Definition of the system}

The energy of any configuration $\gamma\in\Gamma_\Lambda$ or
$\gamma\in\Gamma_0$ is defined by the following formula:
\begin{equation}\label{25}%(2.5)
U_{\phi}(\gamma)=U(\gamma):=\sum_{\{x,y\}\subset\gamma}\phi(|x-y|),
\end{equation}
where $\{\cdot,\cdot\}$ means sum over all possible different
couples of particles from the configuration $\gamma$, $\phi(|x-y|)$-
pair interaction potential.     Define also interaction energy
between configurations $\eta, \,\gamma \in\Gamma_0$ by:
\begin{equation}\label{26}%(2.6)
W(\eta;\gamma):=\sum_{\substack{x\in\eta \\
y\in\gamma}}\phi(|x-y|).
\end{equation}

We introduce 3 kinds of interactions, which will be used in this
article:
\begin{definition}\label{d:1}
Interaction   is called: \\
a)\;stable {\bf (S)}, if there exists $B$>0 such that:\\
\begin{equation}\label{28}%(2.7)
U(\gamma)\geq-B|\gamma|,\; \text{for any\;}\gamma \in \Gamma_{0};
\end{equation}
b)\;superstable {\bf (SS)}, if there exist $A(a)>0, \,B(a) \geq0 $
and  $a>0$ such that:\\
\begin{equation}\label{29}%(2.8)
U(\gamma)\geq A(a) \underset{\Delta\in
\overline{\Delta}_{a}:|\ga_\De|\geq 2 } {\sum}\; |\gamma_{\Delta}|^2
- B(a)|\gamma|,\; \text{for any\;}\;\;\gamma \in \Gamma_{0};
\end{equation}
c)\;strong superstable {\bf (SSS)}, if there exist $A(a)>0,
\,B(a)\geq 0,\;m \geq 2\;$ and   $a_0>0$ such that:\\
\begin{equation}\label{210}%(2.9)
U(\gamma)\geq A(a) \underset{\Delta\in
\overline{\Delta}_{a}:|\ga_\De|\geq 2 } {\sum}\; |\gamma_{\Delta}|^m
- B(a)|\gamma|,\; \text{for any\;}\;\;\gamma \in \Gamma_{0}
\end{equation}
for any $ a \leq a_0$.
\end{definition}
In the above conditions   constants  $A(a),  B(a)$  depend on
$\overline{\Delta}_{a}$ and consequently on $a$. In accordance with
these definitions   there is a  problem to describe the necessary
conditions on  2-body  potential, which ensure stability,
superstability or strong superstability of an infinite statistical
system.  For the latest review and some new results on this problem
see \cite{RT08} and \cite{Te08} for many-body case.

{\bf (A): Assumption on the interaction potential.} \textit{In this
article we consider a general type of potentials \,$\phi$,\, which
are continuous on $\mathbb{R_+}\sm\{0\}$  and for which there exist
\,$ r_0 > 0,\,R > r_0,\,\varphi_{0}>0,\,\varphi_{1}>0,\,
\text{and}\;\, \varepsilon_0
> 0$\, such that:}
\begin{align}
&1)\, \phi(|x|)\equiv -\phi^-(|x|)\geq -
\frac{\varphi_{1}}{|x|^{d+\varepsilon_0}}\;\;\;
  \text{for}\;\;\; |x|\geq R,\label{211};\\%(2.10)
&2)\, \phi(|x|)\equiv\phi^+(|x|)\geq \frac{\varphi_{0}}{|x|^{s}},\,
s\geq d\;\;\;
\text{for}\;\;\; |x|\leq r_0,\label{212}%(2.11)
\end{align}
\textit{where}
\begin{equation}\label{213} %(2.12)
\phi^+(|x|):= \max \{0, \phi(|x|)\}, \, \phi^-(|x|):=-\min \{0,
\phi(|x|)\}.
\end{equation}
Note that  in the Eq. \eqref{210} the constant $a_0\leq r_0$. For
the interaction potentials which satisfy the assumption {\bf (A)}
define two important characteristics (for any
$\Delta\in\overline\Delta_a$ with $a < r_0$ ):
\begin{align}
&1)\quad
 \upsilon_0(a):=\sum_{\Delta'\in\overline\Delta_a}
\;\sup_{x\in\Delta}\;\sup_{y\in\Delta'}\phi^-(|x-y|);\label{214}\\%(2.13)
&2)\quad b(a):=\inf_{\{x,y\}\subset\Delta}\phi^+(|x-y|).\label{215}%(2.14)
\end{align}
Due to the translation invariance of the 2-body potential
$\upsilon_0 $ and $b$ do not depend on the position of $\Delta$. The
following statement is true.

\begin{prop}\label{prop1}
Let potential $\phi$ satisfy the assumption {\bf (A)}. Then the
interaction is strong superstable and the energy $U$ satisfies the
inequality \eqref{210} with some $0<a <a_0$ and if $s>d$ then
\begin{equation}\label{218}%(2.15)
m=2,\;\; A\;=\;A(a)\;=\;\frac{b-2\upsilon_0}{4} >
0,\;\;B\;=\;B(a)\;=\;\frac{\upsilon_0}{2},
\end{equation}
\end{prop}

{\it Proof.} For any $\gamma\in\Gamma_{0}$ and any $a>0$

$$
U(\gamma)=\sum_{\{x,y\}\subset\,\gamma}\phi(|x-y|)=\sum_{\Delta\in\,\overline\Delta_a:|\ga_\De|\geq
2}\sum_{\{x,y\}\subset\,\gamma_\Delta}\phi(|x-y|)
+\sum_{\{\Delta,\Delta'\}\subset\overline\Delta_a}\sum_{\substack {x\in\gamma_\Delta \\
y\in{\gamma_{\Delta'}}}} \phi(|x-y|)
$$
$$
\geq\,\sum_{\Delta\in\,\overline\Delta_a:|\ga_\De|\geq 2}
\frac{1}{2}|\gamma_\Delta|(|\gamma_\Delta|-1) b\,
-\sum_{\{\Delta,\Delta'\}\subset\overline\Delta_a:|\gamma_\Delta|\geq
2,\,|\gamma_{\Delta'}|\geq
2}|\gamma_\Delta||\gamma_{\Delta'}|\,\sup_{x\in\gamma_\Delta}\,\sup_{y\in\gamma_{\Delta'}}\phi^-(|x-y|)
$$
$$
-\;\frac{\upsilon_0}{2}
|\ga|\;\geq\;\sum_{\Delta\in\,\overline\Delta_a:|\ga_\De|\geq 2}
|\gamma_\Delta|^2\left(\frac{
b}{4}-\frac{\upsilon_0}{2}\right)-\frac{\upsilon_0}{2}|\gamma|.
$$
 We use the definitions (\ref{213})--(\ref{215}) and the inequality:
$$
|\gamma_\Delta||\gamma_{\Delta'}|\leq\frac{1}{2}(|\gamma_\Delta|^2+|\gamma_{\Delta'}|^2)
$$
 In the case $s=d$ the  following statement is true (see \cite{RT08}
 for
details): for any sufficiently small $\varepsilon>0$ there exists a
constant $B=B(\varepsilon, a)$ such that the following inequality
holds:
\begin{equation}\label{218_1}
U(\gamma)\geq
\underset{\substack{\Delta\in\overline{\Delta_{a}},\\|\gamma_{\Delta}|\geq
2
}}{\sum}\left(C_{d}\,\log\,|\gamma_{\Delta}|-\frac{v_0}{2}-\varepsilon
\,\log\,|\gamma_{\Delta}| \right)|\gamma_{\Delta}|^{2}- B|\gamma|,
\end{equation}
where (see \cite{HS'06})
\begin{equation}\label{218_2}
C_{d}=\frac{1}{a^{d}}\frac{\pi^{\frac{d}{2}}}{d\,\Gamma\left(\frac{d}{2}\right)}\,\varphi_{0},
\end{equation}
$\Gamma(\cdot)$ is a classical gamma-function.

The system of particles is strong superstable {\bf (SSS)}\, because
for  any $\varepsilon>0$ one can find such numbers $N_0\geq 2$ and
$B=B(N_0; \varepsilon, a)$ that for any $|\gamma_\Delta|>N_0$
\begin{equation}\label{218_3}
C_{d}\,\log\,|\gamma_\Delta|> \frac{v_0}{2}.
\end{equation}
It follows from \eqref{218_1} - \eqref{218_3} that  if $s=d$ we can
put
\[A(a)=K_s(\varepsilon) \upsilon_0, B(a)= L_s (\varepsilon) \upsilon_0+M_s (\varepsilon),\]
where  $K_s(\varepsilon), L_s(\varepsilon), M_s(\varepsilon)$ do not
depend on the parameter  $a$.

In the sequel we will use the estimates \eqref{218} of the constants
$A(a)$ and $B(a)$, because the proof of the main results is the same
for both cases.
$$
\mspace{675mu}\blacksquare
$$
\begin{prop}\label{prop2}
  It follows from the Proposition \eqref{prop1} that for the
  potentials which satisfy the  conditions \eqref{211}- \eqref{213} the
  inequality \eqref{28} holds with
\begin{equation}\label{const_B}
 B=
 \left(\frac{2^{2d-s}d^{\frac{sd}{2}}\phi_0^{s}}{\varphi_0^d}
 \right)^{\frac{1}{s-d}},
\end{equation}
where  the constant $\phi_0$ is very close to $
\underset{\mathbb{R}^d}{\int}\phi^{-}(|x|)dx$.
\end{prop}
 {\it Proof.} We can put $a=a_m$ in such a way that
$b(a_m)=2\upsilon_0(a_m)$. From the definitions \eqref{214},
\eqref{215} it is clear that
\[
b(a_m)\geq \frac{\varphi_0}{d^{\frac{s}{2}}a_m^s}
\]
and $\upsilon_0(a_m)=\frac{1}{a_m^d}\phi_0$ as $\underset{a
\rightarrow 0}{\textnormal{lim}}\; a^d \, \upsilon_0(a)=
\underset{\mathbb{R}^d}{\int}\phi^{-}(|x|)dx$. As a result
\begin{equation}\label{a_m}
a_m \geq\frac{
\bigl(\frac{\varphi_0}{2\phi_0}\bigl)^{\frac{1}{s-d}}}{d^{\frac{s}{2(s-d)}}}.
\end{equation}
The estimate \eqref{const_B} of the constant B directly follows from
\eqref{a_m} and \eqref{218}. The end of proof.
$$
\mspace{675mu}\blacksquare
$$

\begin{remark}\label{B-ndofp}
It is important to stress that the constant $B$ in \eqref{const_B}
does not depend on the partition $\overline{\De}_a$ and depends only
on the potential $\phi$ and dimension of the space.
\end{remark}
\begin{remark}\label{m=1+s/d}
Indeed, for the potentials which satisfy the assumption {\bf (A)}
the inequality \eqref{210} holds with $m=1+s/d$ (see \cite{RT08}).
But for our purpose it is sufficient to apply \eqref{210} with
\eqref{218} and \eqref{28} with \eqref{const_B}.
\end{remark}

\subsection{\bf Partition functions, corresponding pressure and correlation functions}

The main characteristics of Gibbs states are correlation functions.
A family of finite volume correlation functions with empty boundary
conditions for the grand canonical ensemble is defined by the
following formula:
\begin{equation} \label{220}%(2.16)
\rho_\La(\eta;z,\beta) :=\frac{z^{|\eta|}}{Z_\La (z, \beta )}
\int_{\Ga_\La}e^{-\be
U(\eta\cup\ga)}\la_{z\si}(d\ga),\;\;\;\eta\in\Ga_{\La},
\end{equation}
where
\begin{equation}\label{221}%(2.17)
Z_\Lambda(z,\beta ):=\int_{\Gamma_\Lambda}e^{-\beta
U(\gamma)}\lambda_{z\sigma}(d\gamma)
\end{equation}
is the grand partition function which plays the role of normalizing
constant in the definition of the Gibbs measure. Besides it has
independent important physical meaning for the definition of the
thermodynamic function--pressure:
\begin{equation}\label{222}%(2.18)
p(z,\beta) \;=\;\lim_{|\Lambda|\rightarrow \infty}p_\Lambda
(z,\beta)\;=\; \frac{1}{\beta}\lim_{|\Lambda|\rightarrow
\infty}\frac{1}{|\Lambda|}\log Z_{\Lambda}(z, \beta ),
\end{equation}
The existence of this limit for the above defined system of
particles is well-known result (see, {\it e.g.}, \cite{Ru70}).

To define above mentioned approximation let us introduce the
following family of correlation  functions:
\begin{equation} \label{223}%(2.19)
\rho_\La^{(-)}(\eta;z,\beta, a) :=\frac{z^{|\eta|}}{Z_\La^{(-)}(z,
\beta, a)} \int_{\Ga_\La}e^{-\be
U(\eta\cup\ga)}\prod_{\Delta\in\overline{\Delta}_{a}\cap\Lambda}\chi^\Delta_-(\eta\cup\gamma)\la_{z\si}(d\ga),\;\;\;\eta\in\Ga_{\La},
\end{equation}
\begin{equation}\label{224}%(2.20)
 Z_\Lambda^{(-)}(z,\beta,a):=\int_{\Gamma_\Lambda^{dil}}e^{-\beta
U(\gamma)}\lambda_{z\sigma}(d\gamma)=\;
\int_{\Gamma_\Lambda}e^{-\beta
U(\gamma)}\prod_{\Delta\in\overline{\Delta}_{a}\cap\Lambda}\chi^\Delta_-(\gamma)\lambda_{z\sigma}(d\gamma).
\end{equation}
where we introduced $\frak{B}_\De(\Ga_\La)$-measurable function
$\chi_-^{\Delta}$ by the formula:
\begin{equation}\label{225}%(2.21)
  \chi_-^{\Delta}(\gamma) \;
 = \left\{ \begin{array}{ll}
    1,  & \mbox{for $\gamma$ with $N_\De(\ga)=|\gamma_{\Delta}|=0\vee 1,$} \\
    0,  & \mbox{ otherwise}.
           \end{array}
   \right.
\end{equation}

\begin{remark}\label{r1}
By definition  $\rho_\La^{(-)}(\eta; z, \beta; a)=0$ for
$\eta\not\in\Ga_\La^{(dil)}$
\end{remark}

One can define the corresponding pressure:
\begin{equation}\label{226}%(2.22)
p^{(-)}(z,\beta,a ) \;=\;\lim_{|\Lambda|\rightarrow
\infty}p_\Lambda^{(-)} (z,\beta,a)\;=\;
\frac{1}{\beta}\lim_{|\Lambda|\rightarrow
\infty}\frac{1}{|\Lambda|}\log Z_{\Lambda}^{(-)}(z,\beta,a).
\end{equation}

\begin{remark}\label{r2}
The main point of this approximation consists that in expressions
for the basic characteristics of the system integration is carried
out not over all space of configurations $\Ga_\La$, but only over
those configurations which contain for the given partition
$\overline\Delta_a$  not more than one particle in each  cube
$\De\in\overline\Delta_a$. That fact is surprising as for an
infinite system the set of such configurations in $\Ga$ is the set
of  measure zero with respect to the Poisson measure  and the Gibbs
measure. Nevertheless, as we shall see in following section, the
basic characteristics of the approximated system ( even in a
thermodynamic limit $\La\nearrow\R^d$) can be somehow close to the
corresponding characteristics of the initial system.
\end{remark}

\subsection{\bf   Main results }

We prove the results for the infinite volume characteristics, so let
us define the sequence of bounded Lebesgue measurable regions of
$\La_l\subset\R^d$:
\begin{equation}\label{227}%(2.23)
\La_1\subset\La_2\subset\ldots\subset\La_n\subset\ldots,\;\;\underset{l}\cup\La_l\;=\;\R^d.
\end{equation}
 We consider only such  $\Lambda_l \in
\mathcal{B}_c({\R}^{d})$ which is union of cubes $\De_{a}(r)$
defined by \eqref{22}.

\begin{theorem}\label{th1}  Let the interaction potential
$\phi(|x|)$  satisfy  the assumptions {\bf (A)}. Then the limits
\begin{equation}\label{228}%(2.24)
p(z,\beta) \;=\; \frac{1}{\beta}\lim_{l\rightarrow
\infty}\frac{1}{|\Lambda_l|}\log Z_{\Lambda_l}(z, \beta ),
\end{equation}
\begin{equation}\label{229}%(2.25)
p^{(-)}(z,\beta, a ) \;=\; \frac{1}{\beta}\lim_{l\rightarrow
\infty}\frac{1}{|\Lambda_l|}\log Z_{\Lambda_l}^{(-)}(z, \beta, a)
\end{equation}
are finite and for any $\varepsilon
> 0$ there exists \, $a_1 = a_1(z,\varepsilon) > 0$ such that:
\begin{equation}\label{230}%(2.26)
|p(z,\beta) - p^{(-)}(z,\beta,a)| < \varepsilon
\end{equation}
 holds for all positive $z,\;\beta$ and $a \in (0, a_1(z,\varepsilon))$.
\end{theorem}
The proof of the limit \eqref{228} is well known result \cite{Ru70}.
The proof of \eqref{229} and \eqref{230} one can find in
\cite{RT07}. But for the completeness of the presentation we give a
sketch of the proof in the next section.
 A similar result is true
for the correlation functions in the fixed volume $\Lambda$:
\begin{theorem}\label{th_old}  Let the interaction potential
$\phi(|x|)$  satisfy the assumptions {\bf (A)}. Then for any
$\varepsilon
> 0$, any fixed $\Lambda$ and any configuration $\eta\in \Gamma_{0}$  there exists \, $a = a(z, \beta, \varepsilon) > 0$ such that:
\begin{equation}\label{1_old}%(2.26*)
|\rho_\Lambda(\eta;z,\beta) - \rho^{(-)}_\Lambda(\eta;z,\beta, a)|
< \varepsilon.
\end{equation}
\end{theorem}
To formulate a similar result for the limit correlation functions
in  the infinite volume  note that for any configuration
$\eta\in\Ga_0$ and any sequence \eqref{227}, such that
$\eta\subset\La_1$ there exists subsequence $(\La^{'}_{k})$ of
$(\La_{l})$, such that
\begin{equation}\label{231}%(2.27)
\lim_{k\rightarrow \infty}\rho_{\La^{'}_{k}}(\eta;z,\beta)=
\rho(\eta;z,\beta) <\infty
\end{equation}
 for all positive $z,\beta$ uniformly on $\frak{B}_c(\Ga_0)$ . This result follows from the uniform
bounds of the family $\{\rho_\La:\La\in\frak{B}_c(\R^d)\}$. (see
\cite{Ru70}, \cite{Re98}, \cite{PR07}).

It is also clear that the same uniform bounds hold for the family
of $\{\rho^{(-)}_\La:\La\in\frak{B}_c(\R^d)\}$. So, there exists
subsequence ($\La_m^{\prime\prime}$) of the sequence ($\La_k^{'}$) such that one
can define
\begin{equation}\label{231-1}%(2.28)
\rho^{(-)}(\eta;z,\beta, a)\;=\;\lim_{m\rightarrow
\infty}\rho_{\La^{\prime\prime}_{m}}^{(-)}(\eta;z,\beta, a)
 <\infty.
\end{equation}
In the case of small values of a chemical activity z there exists
the unique limit $\rho(\eta; z, \beta)$  that is a solution of
Kirkwood-Salzburg({\bf KS}) equations in the space $E_\xi$ (see
\cite{Ru69}). In the next chapter we will show, that a  similar
equations can be easily written for the functions $\rho^{(-)}(\eta;
z, \beta)$ that is a unique solution of these equations for
sufficiently small values of parameters $z$ or $\beta$.

\begin{theorem}\label{th2}  Let the interaction potential
$\phi(|x|)$  satisfy the assumptions {\bf (A)}. Then for any
$\varepsilon
> 0$, sufficiently small z and any configuration $\eta\in \Gamma_{0}$  there exists \, $a_1 = a_1(z, \beta, \varepsilon) > 0$ such that:
\begin{equation}\label{232}%(2.28)
|\rho(\eta;z,\beta) - \rho^{(-)}(\eta;z,\beta, a)| < \varepsilon.
\end{equation}
 holds for all  $a \in (0, a_1(z,\varepsilon))$.
\end{theorem}
\begin{corollary}\label{co1}
  The inequalities   \eqref {230}, \eqref{1_old},   \eqref {232} ensure the  existence of
 limits:
\begin{equation}\label{233}%(2.29)
\lim_{a\rightarrow 0}p^{(-)}(z,\beta,a)= p(z,\beta)
\end{equation}
 for any positive $z, \beta>0$,  $\eta\in\Gamma_0$;
 \begin{equation}\label{2_old}%(2.29*)
\lim_{a\rightarrow 0}\rho^{(-)}_\Lambda(\eta;z,\beta,a)=
\rho_\Lambda(\eta;z,\beta).
\end{equation}
for any positive $z, \beta>0$,  $\eta\in\Gamma_0$, any fixed
$\Lambda$ and
\begin{equation}\label{234}%(2.30)
\lim_{a\rightarrow 0}\rho^{(-)}(\eta;z,\beta,a)= \rho(\eta;z,\beta).
\end{equation}
 for small positive $z,\textnormal{any} \beta>0$ and   $\eta\in\Gamma_0.$
\end{corollary}

\section{\bf   Proof of Theorem 2.1 }

The proof is based on the expansion which was proposed in
\cite{Re98}. In order to  arrange this expansion let us define also
an indicator of a dense configuration in any cube
$\Delta\in\overline{\Delta}_{a}$ as
\begin{equation*}
\chi_{+}^{\Delta}(\gamma) \;:= \; 1-\chi_{-}^{\Delta}(\gamma).
\end{equation*}
Then we use the following partition of the unity for any
$\gamma\in\Gamma_\Lambda$:

\begin{equation*}%(3.1)
 1 \;= \; \prod_{\Delta\subset \Lambda} \left[
\chi_{-}^{\Delta}(\gamma)+\chi_{+}^{\Delta}(\gamma) \right]
 \;= \; \sum_{n=0}^{N_{\Lambda}}\sum_{\{\Delta_1,...,\Delta_n\}\subset\Lambda}
 \prod_{i=1}^{n}\chi_{+}^{\Delta_{i}}(\gamma)\prod_{\Delta\subset\Lambda\setminus\cup_{i=1}^n\Delta_i}\chi_{-}^{\Delta}(\gamma)
\end{equation*}
\begin{equation}\label{31}%(3.1)
=\;\sum_{ X\subseteq\Lambda}\widetilde\chi_+^{X}(\gamma)
\widetilde\chi_-^{\Lambda\setminus X}(\gamma),
 \end{equation}
where $N_\Lambda\;=\;|\Lambda|/a^d$ (here the symbol $|\cdot|$ means
Lebesgue measure of the set $\Lambda$) is the number of cubes
$\Delta$ in the volume $\Lambda$, and

\begin{equation}\label{32}%(3.2)
\widetilde\chi_{\pm}^{X}(\gamma) \;:= \; \prod_{\Delta\subset
X}\chi_{\pm }^{\Delta}(\gamma).
 \end{equation}
Inserting \eqref{31} into \eqref{221}  we obtain:
\begin{equation}\label{33}%(3.3)
Z_\Lambda(z,\beta)\;=\;\sum_{
X\subseteq\Lambda}\int_{\Gamma_\Lambda}e^{-\beta
U(\gamma)}\,\,\widetilde\chi_+^{X}(\gamma)
\,\widetilde\chi_-^{\Lambda\setminus
X}(\gamma)\lambda_{z\sigma}(d\gamma).
 \end{equation}
It is obvious, that the first term in \eqref{33} (at $X
 =\emptyset$) coincides with $Z_\Lambda^{(-)}(z, \beta, a)$ (see \eqref{224}). Using
infinite divisible property of the Lebesgue-Poisson measure (see for
example (2.5) in \cite{Re93}) one deduce that:
\begin{align}%(3.4)
&Z_\Lambda(z,\beta)\;=\; Z_\Lambda^{(-)}(z, \beta,
a)\left[1+\sum_{\emptyset\ne
X\subseteq\Lambda}\int_{\Gamma_{X}}\w\rho^{(-)}_{\Lambda\setminus X
}(\gamma;a)\,\,\widetilde\chi_+^{ X}(\gamma
)\lambda_{z\sigma}(d\gamma)\right] \notag\\
&:=Z_\Lambda^{(-)}(z,\beta,a)Z_\Lambda^{(+)}(z,\beta,a),\label{34}
 \end{align}
 where
\begin{equation}\label{35}%(3.5)
\w\rho^{(-)}_{\Lambda\setminus X}(\gamma_{X};a)=\frac{e^{-\beta
 U(\gamma_{X})}}{Z_\Lambda^{(-)}(z, \beta, a)}\int_{\Gamma_{\Lambda\setminus
X}}e^{-\beta W(\gamma_{X}\mid\gamma')-\beta U (\gamma')
}\,\,\widetilde\chi_-^{\Lambda\setminus
X}(\gamma')\lambda_{z\sigma}(d\gamma'),
 \end{equation}
We, also, define $p^{(+)}(z,\beta, a)$ in the same way as in
\eqref{229}

\begin{equation}\label{3.6}%(3.6)
p^{(+)}(z,\beta, a ) \;=\; \lim_{l\rightarrow
\infty}p_{\Lambda_l}^{(+)}(z,\beta, a ) \;=\;
\frac{1}{\beta}\lim_{l\rightarrow \infty}\frac{1}{|\Lambda_l|}\log
Z_{\Lambda_l}^{(+)}(z, \beta, a)
\end{equation}

Consequently, in order to prove the Theorem 2.1 we have to estimate
the value of  $p^{(+)}(z,\beta,a).$ Using Proposition 2.1 (Eqs.
\eqref{210}, \eqref{218}) one can obtain:
\begin{equation}\label{37}%(3.7)
e^{-\beta U(\gamma_{X})}\leq \prod_{\Delta\in\overline{\Delta_a}\cap
X}e^{-\beta A(a)|\gamma_\Delta|^2 + \beta B(a)|\gamma_\Delta|},\,
A(a)=\frac{b-2\upsilon_0}{4},\, B(a)= \frac{\upsilon_0}{2}.
\end{equation}
Taking into account  assumption {\bf (A)}(Eqs. \eqref{211}) and
\eqref{214}  we obtain:
\begin{equation}\label{38}%(3.8)
e^{-\beta W(\gamma_{X}| \gamma')}\leq
\prod_{\Delta\in\overline{\Delta_a}\cap X}e^{\beta
\upsilon_0|\gamma_\Delta|}.
\end{equation}
Using infinite divisible property of the measure $\la_{z\si}$ and
using \eqref{37}, \eqref{38} we have:
\begin{align}
&\int_{\Gamma_{X}}\rho^{(-)}_{\Lambda\setminus X
}(\gamma_X;a)\widetilde\chi_+^{X}(\gamma)\lambda_{z\sigma}(d\gamma)\leq
\,\frac{Z_{\Lambda\setminus X }^{(-)}(z,\beta,
a)}{Z_{\Lambda}^{(-)}(z,\beta, a)}\times
\notag\\
&\prod_{\Delta\in\overline{\Delta_a}\cap X}\int_{\Gamma_\Delta}
e^{-\beta A |\gamma_\Delta|^2 + \beta B|\gamma_\Delta|+\beta
\upsilon_0|\gamma_\Delta|}\chi_+^\Delta(\gamma_\Delta)
\lambda_{z\sigma}(d\gamma_\Delta).\notag
 \end{align}
 As a result, using definition of Lebesgue-Poisson measure ( see
\eqref{21})  one can obtain the following estimate:
\begin{equation}\label{39-1}%(3.9)
 \int_{\Gamma_{\Delta}}\,
e^{-\beta\,A\,|\gamma_\Delta|^2+\beta\,(B+\upsilon_0 )
\,|\gamma_\Delta|}\,\chi_+^\Delta(\gamma_\Delta)\,\lambda_{z\sigma}(d\gamma_{\Delta})\,
=\;\sum\limits_{n=2}^{\infty}\,\frac{(a^d\,z)^n}{n!}\,
e^{-\frac{1}{4}\beta\,(b-2\upsilon_0)\,n^2+\frac{3}{2}\beta\,\upsilon_0\,n}\,\leq\,\epsilon_1
(a),
\end{equation}
with
\begin{equation}\label{39}%(3.10)
 \epsilon_1(a)\;=\;\frac{1}{2}\,z^2a^{2d}
\,e^{-\beta\,(b-5\upsilon_0)}\,\, \exp\{z
a^d\,e^{-\beta\,(b-3\upsilon_0)}\}.
\end{equation}
Now from the definition of $N_\Lambda$, $Z_\Lambda^{(+)}(z, \beta,
a)$ (see \eqref{34}) and above estimates we have:
$$
\log Z_\Lambda^{(+)}(z, \beta,
a)\;\leq\;\log\left[1+\sum_{\emptyset\ne
X\subseteq\Lambda}\epsilon_1(a)^{N_{X }}\right]\;=\;
\log\left[1+\sum_{k=1}^{N_\Lambda}\frac{N_\Lambda !}{k!
(N_\Lambda-k)!}\epsilon_1(a)^{k}\right]
$$

\begin{equation}\label{39_1}
\;=\;\log\left[1+\epsilon_1(a)\right]^{N_\Lambda}\;=\;\frac{|\Lambda|}{a^d}\log
\left[1+\epsilon_1(a)\right].
\end{equation}

As a result
$$
p^{(+)}(z,\beta;a)\;\leq\;\frac{1}{\beta a^d}\log
\left[1+\epsilon_1(a)\right].
$$
It is important for the proof of the theorem to find out the
asymptotic behavior of $\epsilon_1(a)$ at $a\rightarrow 0$. It
follows from the Eq.\eqref{39} and the corresponding behavior of $b$
and $\upsilon_0$ (see \eqref{211}-\eqref{215}). As a result we have:
\begin{equation}\label{310}%(3.11)
\epsilon_1(a)\sim a^{2d}e^{-\frac{1}{a^s}},\;\;s\geq d.
\end{equation}
So,
\begin{equation*}
\lim_{a \rightarrow 0}p^{(+)}(z,\beta;a)=0.
\end{equation*}
The end of the proof.
$$
\mspace{675mu}\blacksquare
$$

\section{\bf   Proof of Theorem 2.2 }

Inserting \eqref{31} with an argument $\eta\cup\ga$ into
\eqref{220} we obtain:

\begin{equation}\label{41_old}%(4.1*)
\rho_\Lambda(\eta;z,\beta) \;=
 \; \frac{z^{|\eta|}}{Z_\La(z, \beta)}\sum_{X\subseteq\Lambda}\quad
\int_{\Ga_\La}
   e^{-\beta U(\eta\cup\ga)}\,\widetilde\chi_+^{X}(\eta\cup\ga)
\widetilde\chi_-^{\Lambda\setminus
X}(\eta\cup\ga)\la_{z\si}(d\ga).
\end{equation}
Extracting the first term at $X=\emptyset$ and using the
definitions \eqref{223},\eqref{224} we can rewrite \eqref{41_old}
in the following form:
\begin{equation}\label{42_old}%(4.2*)
\rho_{\Lambda}(\eta;z, \beta) \;= \;\frac{Z_\La^{(-)}(z, \beta, a
)}{Z_\La(z, \beta) }\rho_{\Lambda}^{(-)}(\eta;z, \beta, a)
 + R^{\Lambda}(\eta;z,\beta, a),
\end{equation}
where
\begin{equation}\label{43_old}%(4.3*)
R^{\Lambda}(\eta;z,\beta, a)\;=
 \; \frac{z^{|\eta|}}{Z_\La(z, \beta)}\sum_{\emptyset\ne X\subseteq\Lambda}\quad
\int_{\Ga_\La}
   e^{-\beta U(\eta\cup\ga)}\,\widetilde\chi_+^{X}(\eta\cup\ga)
\widetilde\chi_-^{\Lambda\setminus
X}(\eta\cup\ga)\la_{z\si}(d\ga).
\end{equation}

The proof of the Theorem 2.2 is based on  two  technical lemmas.
\begin{lem}\label{l41}
Let the interaction potential $\phi(|x|)$  satisfy the assumptions
{\bf (A)}. Then for any fixed volume $\La \in\frak{B}_c(\R^d)$ and
any configuration $\eta \in \Gamma_0$ the following holds:
\begin{align}\label{44_old}%(4.4*)
&\underset{a\rightarrow
0}{\textnormal{lim}}R^{\Lambda}(\eta;z,\beta, a)=0.
\end{align}
\end{lem}

{\it Proof.} See  Appendix.\hfill $\blacksquare$
\begin{lem}\label{142}
Let the interaction potential $\phi(|x|)$  satisfy the assumptions
{\bf (A)}.  Then for any fixed volume $ \La\in\frak{B}_c(\R^d)$ the
following holds:
\begin{equation}\label{45_old}
\underset{a\rightarrow
0}{\textnormal{lim}}\frac{Z_\Lambda^{(-)}(z, \beta,
a)}{Z_\Lambda(z, \beta)}=1.
\end{equation}
\end{lem}
{\it Proof.} It follows from the estimates \eqref{39_1}, that
\[\underset{a\rightarrow
0}{\textnormal{lim}}\;\frac{Z_\Lambda^{(-)}(z, \beta,
a)}{Z_\Lambda(z, \beta)}\geq 1.\]

From the other hand in accordance with the definitions \eqref{221},
\eqref{224} it is clear that

\[\frac{Z_\Lambda^{(-)}(z, \beta, a)}{Z_\Lambda(z, \beta)}\leq 1.\]
As a result we have

\[\underset{a\rightarrow
0}{\textnormal{lim}}\frac{Z_\Lambda^{(-)}(z, \beta, a)}{Z_\Lambda(z,
\beta)}=1.\]
\section{\bf   Proof of Theorem 2.3 }
Using  definitions \eqref{21}, \eqref{225}  we can rewrite the
definition \eqref{223} for the family of correlation functions
$\rho_{\Lambda}^{(-)}(\cdot; z, \beta, a)$ in the following form:
\begin{align}\label{40}%4.1
&\rho_{\Lambda}^{(-)}(\eta; z,
\beta,a)=\frac{z^{|\eta|}}{Z_{\Lambda}^{(-)}(z,\beta, a)}e^{-\beta
U(\eta)}\underset{\Delta \in \Lambda_\eta }{\prod}\chi_{-}^\Delta(\eta_\Delta)\biggl[1+\notag \\
&\sum_{k=1}^{N_{\Lambda \setminus \Lambda_\eta}}z^k
\sum_{\{\Delta_1,\ldots,\Delta_k\}\subset (\Lambda\setminus
\Lambda_\eta)\cap \overline{\De}_a}\underset{\Delta_1}{\int} \cdots
\underset{\Delta_k}{\int} e^{-\beta W(\eta;
 \{y_1,\ldots,y_k\})}
e^{-\beta U(\{y_1,\cdots
 y_k\})}dy_1\ldots dy_k \biggl],
\end{align}
where    $\Lambda_\eta$  is  a union of cubes of $\overline{\De}_a$
which contain   points from the configuration $\eta$ (and in the
sequel we will use such a notation) and summation is taken over all
possible sets of cubes from $\overline{\De}_a$ that belong to the
area $\Lambda\setminus\Lambda_\eta$. We prove the theorem using {
\bf KS} equations for the functions $\rho (\eta ; z, \beta)$ and
$\rho^{(-)} (\eta ; z, \beta, a)$. Remind that { \bf KS} equations
for the functions $\rho (\eta ; z, \beta)$ can be written in the
form of one operator equation(see \cite{Ru69})
\begin{equation}\label{41}%4.2
  \rho= z\widetilde{K} \rho+z\delta,
\end{equation}
where operator $\widetilde{K}$ acts on an arbitrary function
$\varphi$ according with the rule
\begin{align}\label{42}%4.3
&(\widetilde{K}\varphi)(\{x_1 \}) =
\sum^{\infty}_{k=1}\frac{1}{k!}\int_{\mathbb{R}^d}\cdots
\int_{\mathbb{R}^d}\prod_{i=1 }^k\left(e^{-\be
\phi(|y_i-x_1|)}-1\right)\times \notag\\
& \varphi(\{y_1,\ldots, y_k\})dy_1 \cdots dy_k,  \textnormal{if }
|\eta|=1\, (\eta= \{x_1 \});
\end{align}

\begin{align}\label{43}%4.4
&(\widetilde{K}\varphi)(\eta)
=\underset{x\in\eta}\sum\widetilde{\pi}(x;\eta\sm\{x\})e^{-\beta
W(x;\eta\sm\{x\})} \bigl[\varphi(\eta \setminus\{x\})+
\sum^{\infty}_{k=1}\frac{1}{k!}\int_{\mathbb{R}^d}\cdots
\int_{\mathbb{R}^d}\times \notag\\
&\prod_{i=1 }^k\left(e^{-\be \phi(|y_i-x|)}-1\right)
\varphi(\eta\sm\{x\}\cup\{y_1,\ldots, y_k\})dy_1 \cdots dy_k \bigl],
\; \textnormal{if } |\eta|\geq 2,
\end{align}
where
\begin{equation}\label{440}%4.5
\widetilde{\pi}(x;\eta\sm\{x\})=
\frac{\pi_W(x;\eta\sm\{x\})}{\underset{y\in\eta}\sum\pi_W(y;\eta\sm\{y\})},\;
\pi_W(x;\eta\sm\{x\})=\begin{cases} 1 &  \text{if $W(x;\eta\sm\{x\})\geq -2B $}, \\
                                    0 &  \text{otherwise},
 \end{cases}
\end{equation}
\begin{equation}\label{44}%4.6
  \rho:=\{\rho(\eta; z, \beta)\}_{\eta \in \Gamma_0},
\end{equation}
 $\delta(\eta)=1 \, \textnormal{if}\,  |\eta|=1 \,
\textnormal{and}\, \delta(\eta)=0 \,\, \textnormal{otherwise}$.

\begin{remark}\label{Ru-notations}
Operator $\widetilde{K}=\Pi K$ in the Ruelle's notation \cite{Ru69}
and \eqref{43}, \eqref{440} are exact realization of the operator
$\Pi$
\end{remark}

Operator $\widetilde{K}$ is bounded operator in  Banach space of
 measurable bounded
 functions  $ E_\xi (\xi>0)$   with the norm
\begin{equation}\label{45}%4.7
  ||\varphi||_\xi= \underset{\eta \in
\Gamma_0}{\textnormal{ sup}}|\varphi(\eta)|\xi^{-|\eta|}
\end{equation}
The solution of the equation \eqref{41} can be represented in the
form of convergent in $E_\xi$ (and point convergent for any fixed
$\eta \in \Gamma_0$ ) series
\begin{equation}\label{46}%4.8
  \rho(\eta;z,\beta)=
\sum_{n=0}^{\infty}z^{n+1} (\widetilde{K}^n \delta)(\eta; z, \beta),
\end{equation}
if

\begin{equation}\label{47}%4.9
 |z|\leq e^{-2\beta B-1}C(\beta)^{-1},\quad
C(\beta)=\int_{\mathbb{R}^d}\left| e^{-\beta\phi(|x|)}-1\right|dx
\end{equation}

and interaction satisfies the conditions \eqref{28}, \eqref{211},
\eqref{212}.

One can write  similar equations for the functions
$\rho_\Lambda^{(-)}(\eta; z,\beta,a)$. It can be easily done  in the
way like it was shown in \cite{Mi} for the case of lattice gas. Let
us proceed with several new notations that correspond the notations
in the space of configurations in the lattice gas system (see
\cite{Mi}). Define the space $C=C_{\overline{\De}_a}$ of
configurations of cubes from $\overline{\De}_a$. Let $s=\{
\Delta_\eta^1,\ldots,\Delta_\eta^{|\eta|}\}$ be the finite
configuration of $|\eta|$ cubes from $\overline{\De}_a$  with all
points from the configuration $\eta \in \Gamma_0$ and $ s^{'}=s
\setminus \{ \Delta_{\eta}^1\}$. Let us denote by
$C_{\overline{\De}_a}^{\textnormal{fin}}$  a space of all finite
configurations of cubes from $C$(see also \cite{Mi}) and
$c=\{\Delta_1,\ldots,\Delta_k\}\in
C_{\overline{\De}_a}^{\textnormal{fin}}$ be any finite configuration
of $k$ cubes from $\overline{\De}_a, \; k=0,\ldots, |\gamma|;
(\textnormal{if}\quad k=0 \quad c= \emptyset)$.

 For technical reason we also introduce new potential
 \[
\hat{\phi}(x, y)=
\phi(|x-y|)+\phi_{\overline{\De}_a}^{\textnormal{cor}}(x,y),
 \]
where
\begin{equation}\label{48}%4.10
 \phi_{\overline{\De}_a}^{\textnormal{cor}}(x,y)= \begin{cases}
    +\infty  & \text{if $x,y \in \Delta \in \overline{\De}_a $}, \\
    0 & \text{if $x \in \Delta, y \in \Delta^{'} $\textnormal{and} \;$  \Delta \neq \Delta^{'}$}.
  \end{cases}
\end{equation}
As in the definition \eqref{40} all points of the configurations
$\eta, \gamma$ are situated in different cubes we can put the
potential $\hat{\phi}$ instead of the potential $\phi$ in the
definitions \eqref{223}, \eqref{224}. Let us define also a potential
$\hat{\phi}(\Delta, \Delta^{'})$ as  the family of potentials
$\hat{\phi}(x, y)$:

\begin{align}\label{49}%4.11
 &\hat{\phi}(\Delta, \Delta^{'})=\left\{  \hat{\phi}(x ,y)\bigl | x \in \Delta, y \in \Delta^{'}
 \right\},\\
&\hat{\phi}(\Delta, \Delta)=+\infty \; \textnormal{for any}\;\Delta
\in \overline{\De}_a. \notag
\end{align}

\begin{remark}
For $c=\{\De_1,...,\De_k\},\;
s=\{\De^1_\eta,...,\De^m_\eta\},\;m=|\eta|$ the functions
$U(c),\;W(s;c)$,\\
$\rho_{\La}^{(-)}(s;z,\beta,a),\;\rho^{(-)}(s;z,\beta,a)$ are the
families (see \eqref{48}) of the corresponding
$U(\ga),\;W(\eta;\ga)$,\\
$\rho_{\La}^{(-)}(\eta;z,\beta,a),\;\rho^{(-)}(\eta;z,\beta,a)$ with
$\ga=\{\ga_{\De_1},...,\ga_{\De_k}\},\;\eta=\{\eta_{\De^1_\eta},...,\eta_{\De^m_\eta}\}$
and at $a\rightarrow 0$ every cube shrinks in the corresponding
point so that $c\rightarrow \ga$,\;$s\rightarrow \eta$.

Configuration $\eta \in \Gamma_0 $ in  definition of the function
  $\rho(\eta; z, \beta)$ is fixed and coordinates of cubes
$\Delta_\eta^1,\ldots,\Delta_\eta^{|\eta|}$ in $\mathbb{R}^d$
change, but  Lebesgue measure of $\La_\eta$ tends to zero
\textnormal{(mess\,$ \Lambda_\eta (a) \rightarrow 0)$}.
\end{remark}

 The energy $U(\gamma)$ of the configuration $\gamma \in
\Gamma_X, \; X\subseteq \Lambda$ in these notations is
\begin{equation}\label{50}%4.12
 U(c)=\underset{1\leq i<j \leq |c|}{\sum}\hat{\phi}(\Delta_i,
 \Delta_j).
\end{equation}
The energy of interaction between configurations of cubes $s,\,c \in
C_{\overline{\De}_a}^{\textnormal{fin}}$ is
\begin{equation}\label{51}%4.13
  W(s;c)=\underset{\Delta \in s,\; \Delta^{'} \in c}{\sum}\hat{\phi}(\Delta,
 \Delta^{'}).
\end{equation}
Then the definition \eqref{40} for the functions
$\rho_\Lambda^{(-)}(\eta; z, \beta, a)$ takes the form;
\begin{equation}\label{52} %4.14
\rho_{\Lambda}^{(-)}(s; z, \beta,a)=
\frac{1}{Z_{\Lambda}^{(-)}(z,\beta, a)}\underset { c \subseteq
 \Lambda \setminus s}{\SI}z^{|s \cup c|}\;e^{-\beta U(s
\cup c)},
\end{equation}
where we introduce new notation
\begin{equation}\label{53}%4.15
 \underset {c \subseteq X}{\SI}f(c)=\sum_{k=0}^{\frac{|X|}{a^d}}\underset{\{\Delta_1,\ldots,\Delta_k\}\subset
\overline{\Delta}_a \cap X}{\sum}\;\;\underset{\Delta_1}{\int}
\cdots \underset{\Delta_k}{\int} f(x_1,\ldots,x_k)dx_1 \cdots dx_k
\end{equation}

Following standard procedure(see \cite{Mi}) one can rewrite
\eqref{52} in the form of Kirkwood-Salzburg equation for the family
of correlation functions $\rho_{\Lambda}^{(-)}(s; z, \beta, a)$:
\begin{align}\label{54}%4.16
&\rho^{(-)}_\La(s;z,\beta, a) = z e^{-\beta W(\Delta_\eta^1;
s^{'})}\biggl\{\rho^{(-)}_\La(s^{'};z,\beta,
a)+ \notag\\
&\underset {\substack{Q \subset
\overline{\De}_a, Q\neq \emptyset\\
Q \cap s =\emptyset }}{\SI}\prod_{\Delta^{'}\in Q}\bigl(e^{-\beta
\hat{\phi}(\Delta_\eta^1; \Delta^{'} )}-1 \bigl)
\rho^{(-)}_\La(s^{'}\cup Q;z,\beta, a)\biggl\}.
\end{align}
Like in the case of functions $\rho_\Lambda$ and  $\rho$ the
equation \eqref{54} can be modified and rewritten in the form of one
operator equation

\begin{equation}\label{55}%4.17
 \rho_\Lambda^{(-)}=z\widetilde{K}_\Lambda^{(-)}\rho_\Lambda^{(-)}+z\delta_\Lambda,
\end{equation}
and for the limit correlation functions $\rho^{(-)}$ we obtain

\begin{equation}\label{56}%4.18
 \rho^{(-)}=z\widetilde{K}^{(-)}\rho^{(-)}+z\delta.
\end{equation}
Operator $\widetilde{K}^{(-)}$ acts on an arbitrary function
$\varphi \in C_{\overline{\De}_a}^{\textnormal{fin}}$ according with
the rule
\begin{equation}\label{57}%4.19
\left(\widetilde{K}^{(-)}\varphi\right)(\{\Delta_\eta^1\};
z,\beta,a) =\underset {Q \subset \overline{\De}_a,  Q\neq
\emptyset}{\SI}\;\underset{\Delta^{'}\in Q}{\prod}\bigl(e^{-\beta
\hat{\phi}(\Delta_\eta^1; \Delta^{'} )}-1 \bigl) \varphi( Q;z,\beta,
a)
\end{equation}
for $|s|=1$, and
\begin{align}\label{58}%4.20
&\left({\widetilde K}^{(-)}\varphi\right)(s; z, \beta, a)
=\underset{\De\in s}\sum{\widetilde\pi}(\De;s') e^{-\beta
W(\Delta; s^{'})}\biggl\{\varphi(s^{'})+\\
&\underset {\substack{Q \subset
\overline{\De}_a, Q\neq \emptyset\\
Q \cap s =\emptyset }}{\SI}\underset{\Delta^{'}\in
Q}{\prod}\bigl(e^{-\beta \hat{\phi}(\Delta, \Delta^{'} )}-1 \bigl)
\varphi(s^{'}\cup Q)\biggl\} \;\;\textnormal{for} \;|s|\geq 2.
\notag
\end{align}

Proof of existence of the solutions of the equations \eqref{55},
\eqref{56} in the form of convergent series
\begin{equation}\label{59}%4.21
  \rho^{(-)}_\La(\cdot;z,\beta, a)=
\sum_{n=0}^{\infty}z^{n+1}\left(\left(\widetilde{K}_{\Lambda}^{(-)}
\right)^n\delta\right)(\cdot; z, \beta, a),
\end{equation}

\begin{equation}\label{60}%4.22
  \rho^{(-)}(\cdot;z,\beta, a)=
\sum_{n=0}^{\infty}z^{n+1}\left(\left(\widetilde{K}^{(-)}
\right)^n\delta\right)(\cdot; z, \beta, a)
\end{equation}
and the equality

\begin{equation}\label{61}%4.23
 \underset{\Lambda \nearrow \mathbb{R}^d}{\textnormal{lim}}
 \;\rho_\Lambda^{(-)}(s; z, \beta, a)= \rho^{(-)}(s; z, \beta, a),\;\; s \in C_{\overline{\De}_a}^{\textnormal{fin}}
\end{equation}
for $z, \beta$ that yield the conditions \eqref{47} can be done
 in a similar way as in work \cite{Mi}.

 So, we have to show that the solution \eqref{60} of the equation \eqref{56}
 converges to the solution of  Kirkwood-Salzburg equation \eqref{41} if $a\rightarrow
 0$.

In the sequel in the expressions for the operators $\widetilde{K},
\widetilde{K}^{(-)} $ we will consider only the case $|s|\geq 2$, as
the case $|s|=1$ is rather similar.

Due to the convergence of the series \eqref{59},\eqref{60} uniformly
in $a$ it is sufficient to prove the point convergence
$(\widetilde{K}^{(-)}\delta)^n \rightarrow (\widetilde{K} \delta)^n$
for
 any $n\geq 1$. It implies
 obviously $\rho^{(-)}(\cdot; z,\beta,a)\rightarrow \rho(\cdot; z,\beta,
 a)$ if $a \rightarrow 0$ for sufficiently small values of a
 chemical activity z. To prove this statement let us use  method of mathematical induction.  Let us put $n=1$ (base of induction). We
 have from \eqref{42}, \eqref{43}, \eqref{57}, \eqref{58}:
\[
(\widetilde{K}^{(-)}\delta)(s)=
\begin{cases}
\underset{\mathbb{R}^d\setminus \De^1_\eta}{\int}\bigl(e^{-\beta
\phi(|y-x_1|)}-1\bigl)dy
&\text{if $|s|=1$,}\\
\left(\underset{\Delta \in
\overline{\Delta}_a}{\prod}\chi_{-}^{\Delta}(\eta)\right)e^{-\beta\phi(|x_2-x_1|)} &\text{if $|s|=2$,}\\
0 &\text{if $|s|>2$;}
\end{cases}
\]
\[
(\widetilde{K} \delta)(\eta)=
\begin{cases}
\underset{\mathbb{R}^d}{\int}\bigl(e^{-\beta
\phi(|y-x_1|)}-1\bigl)dy
&\text{if $|\eta|=1$,}\\
e^{-\beta\phi(|x_2-x_1|)} &\text{if $|\eta|=2$,}\\
0 &\text{if $|\eta|>2$.}
\end{cases}
\]
It is clear that $\widetilde{K}^{(-)}\delta \rightarrow
\widetilde{K} \delta \,\, \textnormal{if}\,\, a \rightarrow 0 $ in
the sense of point convergence. It is useful to notice that
$\left((\widetilde{K}^{(-)})^n\delta\right)(s) =
\left((\widetilde{K})^n\delta\right)(\eta) =0 \quad \textnormal{if}
\,\,|s|>n+1$ ( $|\eta|>n+1$). Let us make the step of induction. Let
$(\widetilde{K}^{(-)})^n\delta \rightarrow (\widetilde{K})^n\delta$
in the sense of point convergence. Using this assumption we have to
prove that $(\widetilde{K}^{(-)})^{n+1}\delta \rightarrow
(\widetilde{K} )^{n+1}\delta$ in the same sense. It follows from
\eqref{43}, \eqref{58} that ($|\eta|=|s|\geq 2$)

\begin{align}\label{62}%4.24
 &((\widetilde{K}^{(-)})^{n+1}\delta)(s)=\underset{\De\in s}\sum\widetilde{\pi}(\De;s')e^{-\beta
W(\Delta; s^{'})} \underset{\Delta \in
\overline{\Delta}_a}{\prod}\chi
_{-}^{\Delta}(\eta)\biggl\{((\widetilde{K}^{(-)})^{n}\delta)(s^{'})+\notag \\
&\sum_{k=1}^{n-|s|+2}\underset {\substack{Q \subset
\overline{\De}_a, Q\neq \emptyset\\
Q \cap s =\emptyset, |Q|=k}}{\SI}\underset{\Delta^{'}\in
Q}{\prod}\bigl(e^{-\beta \hat{\phi}(\Delta \Delta^{'} )}-1 \bigl)
((\widetilde{K}^{(-)})^{n}\delta)(s^{'}\cup Q)\biggl\},
\end{align}
\begin{align}\label{63}%4.25
&(\widetilde{K}^{n+1}\delta)(\eta)=\underset{x\in\eta}\sum\widetilde{\pi}(x;\eta\sm\{x\})e^{-\beta
W(\{x_1\}; \eta \setminus \{x_1\} } \biggl\{(K^{n}\delta)(\eta
\setminus
\{x_1\})+ \notag \\
&\sum_{k=1}^{n-|\eta|+2}\frac{1}{k!}\underset{(\mathbb{R}^d)^k}{\int}
\underset{1\leq i\leq k}{\prod} \bigl(
e^{-\beta\phi(|y_i-x_1|)}-1\bigl)(K^{n}\delta)(\eta \setminus
\{x_1\}\cup \{y_1,\ldots, y_k\} )dy_1\cdots dy_k\biggl\}.
\end{align}
Note that $(\widetilde{K}^{n}\delta)(\eta \setminus \{x_1\}\cup
\{y_1,\ldots, y_k\})\, \textnormal{and}
\;((\widetilde{K}^{(-)})^{n}\delta)(s^{'}\cup Q)$ are  measurable
bounded functions as operators $\widetilde{K}, \widetilde{K}^{(-)}$
are bounded in the spaces $E_\xi$ with some $\xi>0$. Besides because
of stability condition \eqref{28}: $\underset{1\leq i \leq
k}{\prod}\left(e^{-\beta\phi(|y_i-x_1|)}-1 \right)\leq \left|
e^{2\beta B}-1\right|^k< +\infty$. Then the proof of the theorem is
based on one technical lemma.
\begin{lem}\label{lem_last}
Let $F_{-}(\cdot; a),\;F(\cdot)\in L^1(\R^{dk})$ be  symmetric
bounded functions of its  variables,
 and $\underset{a\rightarrow
0}{\textnormal{lim}}\,F_{-}(x_1,\ldots,x_k; a)=F(x_1,\ldots,x_k)$
for any $(x_1,\ldots,x_k)\in (\mathbb{R}^d)^k$. Then the following
equality is true:
\begin{equation}\label{64}%4.25
\underset{a \rightarrow
0}{\textnormal{lim}}\underset{\{\Delta_1,\ldots,\Delta_k\}\subset
\overline{\Delta}_a \setminus
\Lambda_\eta}{\sum}\;\underset{\Delta_1}{\int}dx_1 \cdots
\underset{\Delta_k}{\int}dx_k F_{-}(x_1,\ldots,x_k;
a)=\frac{1}{k!}\underset{(\mathbb{R}^d
)^k}{\int}F(x_1,\ldots,x_k)dx_1\cdots dx_k.
\end{equation}
Proof

See Appendix.
\end{lem}
The step of induction follows directly from \eqref{53} \eqref{62},
\eqref{63} and the statement of the lemma. Theorem is proven.
$$
\mspace{675mu}\blacksquare
$$
\section{\bf   Appendix }

{\it Proof of the  lemma \ref{l41}}

 One can rewrite
\eqref{43_old}in such a  way:
\begin{align}\label{A3}%(A.3)
 &R^{\Lambda}(\eta;z,\beta, a)= \;\frac{z^{|\eta|}}{Z_\La(z,\beta)}
 \underset{\emptyset \neq X \subseteq \Lambda \cap
 \overline{\Delta}_a}{\sum}\int_{\Ga_\La}e^{-\beta U(\eta\cup
 \gamma_X)}\widetilde\chi_+^{X}(\eta\cup\ga) e^{-\beta W(\eta \cup \gamma_X; \gamma_{\Lambda\setminus
 X})}\times \notag\\
 &e^{-\beta U(\gamma_{\Lambda\setminus X})}\widetilde\chi_-^{\Lambda \setminus X}(\eta\cup\ga)\la_{z\si}(d\ga).
\end{align}
Using infinite divisibility property of Lebesgue-Poisson  measure,
 an  obvious estimate: $ e^{-\beta W(\eta \cup \gamma_X;
\gamma_{\Lambda\setminus
 X})}\leq e^{\beta\upsilon_0(|\eta|+|\gamma_X|)}$  and the fact that $\widetilde \chi_-^{\Lambda \setminus X}(\eta \cup\ga)\leq 1$, we obtain from
 \eqref{A3}:
\begin{align}\label{A4}%(A.4)
 &R^{\Lambda}(\eta;z,\beta, a)\leq \;\frac{({z e^{\beta\upsilon_0})^{|\eta|}}}{Z_\La(z,\beta)}
 \underset{\emptyset \neq X \subseteq \Lambda \cap
 \overline{\Delta}_a}{\sum}\int_{\Ga_X}e^{-\beta (U(\eta\cup
 \gamma_X)+\upsilon_0 |\gamma_X|)}\widetilde\chi_+^{X}(\eta\cup\ga)\la_{z\si}(d\ga_{ X}) \times \notag\\
 &\int_{\Gamma_{\Lambda\setminus X}}e^{-\beta U(\gamma_{\Lambda\setminus X})}\la_{z\si}(d\ga_{\Lambda\setminus X}).
\end{align}
Let us take into account that $Z_{\Lambda\setminus X }(z, \beta
)=\int_{\Gamma_{\Lambda\setminus X}}e^{-\beta
U(\gamma_{\Lambda\setminus X})}\la_{z\si}(d\ga_{\Lambda\setminus
X})$
 and \\
$Z_{\Lambda\setminus X }(z, \beta )\leq Z_{\Lambda }(z, \beta )$.
Then we have from \eqref{A4}:
\begin{equation}\label{A5}%(A.5)
 R^{\Lambda}(\eta;z,\beta, a)\leq \;({z e^{\beta\upsilon_0})^{|\eta|}}
 \underset{\emptyset \neq X \subseteq \Lambda \cap
 \overline{\Delta}_a}{\sum}\int_{\Ga_X}e^{-\beta (U(\eta\cup
 \gamma_X)+\upsilon_0 |\gamma_X|)}\widetilde\chi_+^{X}(\eta\cup\ga)\la_{z\si}(d\ga_{
 X}).
\end{equation}
Using Proposition 2.1 we deduce that the interaction is superstable
with the constants $A,B$, that are taken from \eqref{218} and
\begin{align}\label{A6}%(A.6)
 &R^{\Lambda}(\eta;z,\beta, a)\leq \;({z e^{\beta(\upsilon_0+B)})^{|\eta|}}
 \underset{\emptyset \neq X \subseteq  \overline{\Delta}_a\cap\Lambda
}{\sum}\int_{\Ga_X}e^{ \underset{\Delta\in (X \cup
\Lambda_\eta)\cap\overline{\Delta}_{a}: |\ga_\De|+|\eta_\Delta|\geq
2 } {\sum}\;\beta \left(-A(|\eta_\Delta|+ |\gamma_{\Delta}|)^2 +
(B+\upsilon_0)
|\gamma_\Delta|\right)}\times\\
&\widetilde\chi_+^{X}(\eta\cup\ga)\la_{z\si}(d\ga_{
 X})=({z e^{\beta(\upsilon_0+B)})^{|\eta|}}( R_1^\Lambda+R_2^\Lambda),\notag
\end{align}
where
\begin{align}\label{A61}
&R_1^\Lambda= \underset{\emptyset \neq X \subseteq
\overline{\Delta}_a\cap (\Lambda
\setminus\Lambda_\eta)}{\sum}\int_{\Ga_X}e^{ \underset{\Delta\in X
\cap\overline{\Delta}_{a}: |\ga_\De|\geq 2 } {\sum}\; \beta
\left(-A|\gamma_{\Delta}|^2 + (B+\upsilon_0)
|\gamma_\Delta|\right)}\times\\
&\widetilde\chi_+^{X}(\eta\cup\ga)\la_{z\si}(d\ga_{
 X}),\notag
\end{align}
\begin{align}\label{A62}
&R_2^\Lambda= \underset{ \substack{\emptyset \neq X \subseteq
\overline{\Delta}_a\cap \Lambda,\\X\cap\Lambda_\eta\neq \emptyset}
}{\sum}\int_{\Ga_X}e^{ \underset{\Delta\in (X \cup
\Lambda_\eta)\cap\overline{\Delta}_{a}: |\ga_\De|+|\eta_\Delta|\geq
2 } {\sum}\; \beta \left(-A(|\gamma_{\Delta}|+|\eta_\Delta|)^2 +
(B+\upsilon_0)
|\gamma_\Delta|\right)}\times\\
&\widetilde\chi_+^{X}(\eta\cup\ga)\la_{z\si}(d\ga_{
 X}).\notag
\end{align}

Note, that if $\Lambda \subseteq\Lambda_\eta $ then $R_1^\Lambda=0$.
Using the same technique as in \eqref{39-1}- \eqref{310} and putting
in \eqref{A61} $\Lambda \setminus\Lambda_\eta\neq \emptyset$  we
obtain:
\begin{align}\label{A63}
&R_1^\Lambda \leq \;
(1+\epsilon_1(a))^{\frac{|\Lambda\setminus\Lambda_\eta|}{a^d}}-1\leq
\epsilon_1(a)\frac{|\Lambda\setminus\Lambda_\eta|}{a^d}
(1+\epsilon_1(a))^{\frac{|\Lambda\setminus\Lambda_\eta|}{a^d}-1}.
\end{align}

 Using once again infinite divisibility property of
Lebesgue-Poisson  measure we have from \eqref{A62}:
\begin{align}\label{A7}%(A.7)
 &R_2^\Lambda =\;
 \underset{ \substack{\emptyset \neq X \subseteq
\overline{\Delta}_a\cap \Lambda,\\X\cap\Lambda_\eta\neq \emptyset}
}{\sum}R_0^\Lambda(\eta_{X \cap \Lambda_\eta};z,\beta, a)\times \notag\\
 &\int_{\Ga_{X \setminus \Lambda_\eta}}e^{ \underset{\Delta\in
\overline{\Delta}_{a}\cap(X \setminus \Lambda_\eta): |\ga_\De|\geq 2
} {\sum}\; \beta \left(-A|\gamma_{\Delta}|^2 + (B+\upsilon_0)
|\gamma_\Delta|\right)} \widetilde\chi_+^{X
\setminus\Lambda_\eta}(\ga_{X\setminus \Lambda_\eta
})\la_{z\si}(d\ga_{
 X\setminus\Lambda_\eta}),
\end{align}
where
\begin{align}\label{A8}%A8
  &R_0^\Lambda(\eta;z,\beta, a)=\int_{\Ga_{X\cap \Lambda_\eta}}e^{ \underset{\Delta\in
\overline{\Delta}_{a}\cap X \cap \Lambda_\eta:
|\ga_\De|+|\eta_\Delta|\geq 2 } {\sum}\;\beta
\left(-A(|\eta_{\Delta}|+|\gamma_{\Delta}|)^2 + (B+\upsilon_0)
|\gamma_\Delta|\right)}\times \notag\\
&\widetilde\chi_+^{X\cap \Lambda_\eta}(\eta\cup\ga_{X\cap
\Lambda_\eta })\la_{z\si}(d\ga_{
 X\cap \Lambda_\eta})=
 \underset{\Delta \in \overline{\Delta}_a \cap X \cap \Lambda_\eta}{\prod}\int_{\Gamma_\Delta}e^{\beta \left(-A (|\eta_{\Delta}|+|\gamma_{\Delta}|)^2 + (B+\upsilon_0)
|\gamma_\Delta|\right)}\times\notag\\
&\chi_+^{\Delta}(\eta_\Delta\cup\gamma_\Delta)\la_{z\si}(d\ga_{
 \Delta})
 \leq \underset{\Delta \in \overline{\Delta}_a \cap X \cap \Lambda_\eta}{\prod}\int_{\Gamma_\Delta}e^{\beta \left(-A |\gamma_{\Delta}|^2 + (B+\upsilon_0)
|\gamma_\Delta|\right)} \la_{z\si}(d\ga_{
 \Delta})
\end{align}
Substituting values of the constants $A,B$ into \eqref{A8} one can
obtain such an estimate:
\begin{equation}\label{A9}%A9
R_0^\Lambda(\eta;z,\beta, a)\leq
e^{-\beta\left(\frac{b}{4}-2\upsilon_0\right)}\underset{\Delta \in
\overline{\Delta}_a \cap X \cap
\Lambda_\eta}{\prod}\int_{\Gamma_\Delta} \la_{z\si}(d\ga_{
 \Delta})\leq e^{-\beta\left(\frac{b}{4}-2\upsilon_0\right)}e^{z a^d |\eta|}
\end{equation}

Using \eqref{A7}, \eqref{A9}  we can  estimate $R_2^\Lambda$ from
above in the form:
\begin{align}\label{A10}%(A.10)
 &R_2^{\Lambda}\leq e^{-\beta\left(\frac{b}{4}-2\upsilon_0\right)}e^{z a^d |\eta|}
 \underset{ \substack{\emptyset \neq X \subseteq
\overline{\Delta}_a\cap \Lambda,\\X\cap\Lambda_\eta\neq \emptyset}
}{\sum}\;\;\;\int_{\Ga_{X \setminus \Lambda_\eta}}e^{
\underset{\Delta\in \overline{\Delta}_{a}\cap(X \setminus
\Lambda_\eta): |\ga_\De|\geq 2 } {\sum}\; \beta
\left(-A|\gamma_{\Delta}|^2 + (B+\upsilon_0)
|\gamma_\Delta|\right)}\times \notag\\
&\widetilde\chi_+^{X \setminus\Lambda_\eta}(\ga_{X\setminus
\Lambda_\eta })\la_{z\si}(d\ga_{
 X\setminus\Lambda_\eta}).
\end{align}
Let us take into account that for any $\frak{B}(\Ga_\La)$-measurable
function $F(\gamma)$ the following holds:
\[
\underset{ \substack{\emptyset \neq X \subseteq
\overline{\Delta}_a\cap \Lambda,\\X\cap\Lambda_\eta\neq \emptyset}
}{\sum}\;\;\;\int_{\Ga_{X \setminus
\Lambda_\eta}}F(\gamma_{X\setminus \Lambda_\eta})\la_{z\si}(d\ga_{
 X\setminus \Lambda_\eta})\leq(2^{|\eta|}-1)
\underset{ X \subseteq \overline{\Delta}_a\cap \Lambda \setminus
\Lambda_\eta  }{\sum}\;\;\;\int_{\Ga_{X
}}F(\gamma_X)\la_{z\si}(d\ga_{
 X})
\]
Using this fact and infinite divisibility property of
Lebesgue-Poisson measure we obtain from \eqref{A10}:
\begin{align}\label{A11}%(A.11)
 &R_2^{\Lambda}\leq e^{-\beta\left(\frac{b}{4}-2\upsilon_0\right)}e^{a^d |\eta|}
 (2^{|\eta|}-1)
\underset{ X \subseteq \overline{\Delta}_a\cap \Lambda \setminus
\Lambda_\eta  }{\sum}\;\underset{\Delta \in
X}{\prod}\;\int_{\Gamma_\Delta}e^{\beta \left(-A |\gamma_{\Delta}|^2
+ (B+\upsilon_0) |\gamma_\Delta|\right)}\times \notag
\\
&\chi_+^{\Delta}(\ga_{\Delta})\la_{z\si}(d\ga_{
 \Delta})\leq e^{-\beta\left(\frac{b}{4}-2\upsilon_0\right)}e^{z a^d |\eta|}
 (2^{|\eta|}-1) \left(1+\epsilon_1(a) \right)^{\frac{|\Lambda\setminus\Lambda_\eta|}{a^d}}.
\end{align}
It follows from \eqref{A6}, \eqref{A63}, \eqref{A11} that:
\begin{align}\label{A12}%(A12)
&R^{\Lambda}(\eta;z,\beta, a)\leq (z e^{\beta
\frac{3}{2}\upsilon_0})^{|\eta|}\left(1+\epsilon_1(a)\right)^{\frac{|\Lambda\setminus
\Lambda_\eta
|}{a^d}-1}\biggl(\epsilon_1(a)\frac{|\Lambda\setminus\Lambda_\eta|}{a^d}+(2^{|\eta|}-1)\times\notag\\
&(1+\epsilon_1(a))e^{-\beta\left(\frac{b}{4}-2\upsilon_0\right)}e^{z
a^d |\eta|}\biggl)\rightarrow 0, \textnormal{if}\; a \rightarrow 0.
\end{align}

The lemma is proven.
$$
\mspace{675mu}\blacksquare
$$

{\it  Proof of the  lemma \eqref{lem_last}}

We have to prove that for any $\varepsilon>0$ there exists
$a_\varepsilon$ that for any $a<a_\varepsilon$ the following
estimate holds:

\begin{equation}\label{65}%5.1
\left|  \underset{\{\Delta_1,\ldots,\Delta_k\}\subset
\overline{\Delta}_a \setminus
\Lambda_\eta}{\sum}\;\underset{\Delta_1}{\int}dx_1 \cdots
\underset{\Delta_k}{\int}dx_k F_{-}(x_1,\ldots,x_k;
a)-\frac{1}{k!}\underset{(\mathbb{R}^d
)^k}{\int}F(x_1,\ldots,x_k)dx_1\cdots dx_k\right|<\varepsilon.
\end{equation}

From the integrability conditions of the functions $F_{-},\, F$ one
can obtain that for any $\varepsilon>0$ there exists bounded
$\Lambda_\varepsilon\subset\R^d$, such that
\begin{align}\label{66}%5.2
&\biggl|  \underset{\{\Delta_1,\ldots,\Delta_k\}\subset
\overline{\Delta}_a \setminus
\Lambda_\eta}{\sum}\;\underset{\Delta_1}{\int}dx_1 \cdots
\underset{\Delta_k}{\int}dx_k F_{-}(x_1,\ldots,x_k; a)- \notag \\
&\underset{\{\Delta_1,\ldots,\Delta_k\}\subset
(\overline{\Delta}_a\setminus\Lambda_\eta)\cap\Lambda_\varepsilon
}{\sum}\;\underset{\Delta_1}{\int}dx_1 \cdots
\underset{\Delta_k}{\int}dx_k F_{-}(x_1,\ldots,x_k;
a)\biggl|<\frac{\varepsilon}{3},
\end{align}
and
\begin{equation}\label{67}%5.3
\biggl|
\frac{1}{k!}\underset{(\mathbb{R}^d)^k}{\int}F(x_1,\ldots,x_k)dx_1\cdots
dx_k-\frac{1}{k!}\underset{\Lambda_\varepsilon^k}{\int}F(x_1,\ldots,x_k)dx_1\cdots
dx_k\biggl|<\frac{\varepsilon}{3}.
\end{equation}
Using \eqref{65}-- \eqref{67} it is easy to notice that the proof of
the lemma can be reduced to verification the fact that for any
$\varepsilon>0$ there exists $a_\varepsilon=f(\varepsilon)>0$ that
for any  $a<a_\varepsilon$ the following estimate is true:
\begin{equation}\label{68}%5.4
R=\biggl|  \underset{\{\Delta_1,\ldots,\Delta_k\}\subset
(\overline{\Delta}_a\setminus\Lambda_\eta)\cap\Lambda_\varepsilon
}{\sum}\;\underset{\Delta_1}{\int}dx_1 \cdots
\underset{\Delta_k}{\int}dx_k F_{-}(x_1,\ldots,x_k;
a)-\frac{1}{k!}\underset{\Lambda_\varepsilon^k}{\int}F(x_1,\ldots,x_k)dx_1\cdots
dx_k\biggl|<\frac{\varepsilon}{3}.
\end{equation}
Dividing each integral over $\La_\varepsilon$ into the sum of
integrals over $\De\in\overline{\De}_a\cap\La_\varepsilon$ one can
arrange two terms in \eqref{68} into three ones to get estimate

 $$R\leq R_1+R_2+R_3,$$

\begin{align}\label{69}%5.5
&R_1=  \sum_{j=1}^{k-1}\,\,\, \underset{\substack{ \{k_1,\ldots,k_j\},\\
 k_1+\cdots+k_j=k}}{\sum} \frac{1}{k_1! \cdots k_j!}\underset{\pi \in P_j}{\sum^{'}}\;\;\underset{\{\Delta_1,\ldots,\Delta_j\}\subset \overline{\Delta}_a\cap\Lambda_\varepsilon}{\sum}\notag\\
 &\underset{\Delta_1}{\int}dx_1\cdots
 \underset{\Delta_1}{\int}dx_{k_{\pi(1)}}\cdots
 \underset{\Delta_{j}}{\int}dx_{k-k_{\pi(j)}+1}\cdots
 \underset{\Delta_{j}}{\int}
 |F(x_1,\ldots,x_k)| dx_k ,
\end{align}
\begin{equation}\label{70}%5.6
R_2 =  \underset{\{\Delta_1,\ldots,\Delta_k\}\subset
(\overline{\Delta}_a\sm\La_\eta)\cap\Lambda_\varepsilon
}{\sum}\;\underset{\Delta_1}{\int}dx_1 \cdots
\underset{\Delta_k}{\int}dx_k |F_{-}(x_1,\ldots,x_k;
a)-F(x_1,\ldots,x_k)|,
\end{equation}
\begin{equation}\label{71}%5.7
R_3=\underset{\substack{\{\Delta_1,\ldots,\Delta_k\}\subset
\overline{\Delta}_a \cap\Lambda_\varepsilon,\\
\{\Delta_1,\ldots,\Delta_k\}\cap\Lambda_\eta\neq \emptyset
}}{\sum}\;\underset{\Delta_1}{\int}dx_1 \cdots
\underset{\Delta_k}{\int}dx_k |F(x_1,\ldots,x_k)|,
\end{equation}

where $P_j$ is a set of all permutations of numbers $\{1,\ldots,j
\}$, but the sum $\underset{\pi \in P_j}{\sum^{'}}$ means that we
consider only different permutations of numbers
$\{k_1,\ldots,k_j\}$(for example if $k_i=k_j$ the permutation of
numbers $k_i, k_j$ is considered only once). Then for $R_1$ we have:
\begin{align}
& R_1 <  \sum_{j=1}^{k-1}\frac{1}{j!}\, \underset{\substack{ \{k_1,\ldots,k_j\}, \\
 k_1+\cdots+k_j=k}}{\sum}\frac{1}{k_1! \cdots k_j!}\,\, \underset{\pi \in P_j}{\sum^{'}} \;\;\underset{\Delta_1 \subset \overline{\Delta}_a\cap\Lambda_\varepsilon,\ldots, \Delta_j \subset \overline{\Delta}_a\cap\Lambda_\varepsilon}{\sum}\notag\\
 &\underset{\Delta_1}{\int}dx_1\cdots
 \underset{\Delta_1}{\int}dx_{k_{\pi(1)}}\cdots
 \underset{\Delta_{j}}{\int}dx_{k-k_{\pi(j)}+1}\cdots
 \underset{\Delta_{j}}{\int}
 |F(x_1,\ldots,x_k)| dx_k  \notag<
\end{align}
\begin{equation*}
 \sum_{j=1}^{k-1}\, \frac{a^{dk}}{j!}\,\;\underset{\substack{
\{k_1,\ldots,k_j\},\\
 k_1+\cdots+k_j=k}}{\sum}\frac{1}{k_1! \cdots k_j!}\,\, \underset{\pi \in P_j}{\sum^{'}}\;\;\underset{\Delta_1\subset
\overline{\Delta}_a\cap\Lambda_\varepsilon, \ldots , \Delta_j\subset
\overline{\Delta}_a\cap\Lambda_\varepsilon}{\sum}
  \,\underset{\{x_1,\ldots,x_k\}\in (\mathbb{R}^d)^k}{\textnormal{sup}}\,
 |F(x_1,\ldots,x_k)|<
\end{equation*}

\begin{equation}\label{72}%5.8
 \sum_{j=1}^{k-1}\,
\frac{a^{d(k-j)}}{j!}\,|\Lambda_\varepsilon|^j\,
\underset{\substack{
\{k_1,\ldots,k_j\},\\
 k_1+\cdots+k_j=k}}{\sum}\frac{1}{k_1! \cdots k_j!}\,\, \underset{\pi \in P_j}{\sum^{'}}
  \,\underset{\{x_1,\ldots,x_k\}\in (\mathbb{R}^d)^k}{\textnormal{sup}}\,
 |F(x_1,\ldots,x_k)|\rightarrow 0 \,\, \textnormal{if}\,\,
 a\rightarrow 0.
\end{equation}
For $R_2$:

\begin{align}\label{73}%5.9
&R_2< \frac{1}{k!}\underset{\Delta_1\subset
\overline{\Delta}_a\cap\Lambda_\varepsilon, \ldots , \Delta_k\subset
\overline{\Delta}_a\cap\Lambda_\varepsilon}{\sum}\;\underset{\Delta_1}{\int}dx_1
\cdots \underset{\Delta_k}{\int}dx_k \,\, |F_{-}(x_1,\ldots,x_k;
a)-F(x_1,\ldots,x_k)|<\notag \\
& \frac{|\Lambda_\varepsilon|^k}{k!} \underset{\{x_1,\ldots,x_k\}\in
(\mathbb{R}^d)^k}{\textnormal{sup}}|F_{-}(x_1,\ldots,x_k;
a)-F(x_1,\ldots,x_k)|\rightarrow 0\, \textnormal{if}\, a\rightarrow
0,
\end{align}
and for $R_3$:
\begin{align}\label{74}%5.10
&R_3= \sum_{i=1}^{\textnormal{min}(|\eta|; k)
}\underset{\{\Delta_1,\ldots,\Delta_i\}\subset \Lambda_\eta
}{\sum}\;\underset{\{\Delta_{i+1},\ldots,\Delta_k\}\subset
\overline{\Delta_a}\cap\Lambda_\varepsilon \setminus \Lambda_\eta
}{\sum}\;\;\underset{\Delta_1}{\int}dx_1 \cdots
\underset{\Delta_k}{\int}dx_k \,\, |F(x_1,\ldots,x_k)|< \notag \\
& \sum_{i=1}^{\textnormal{min}(|\eta|; k) }\frac{|\La_\eta|^i}{i!}
\;\frac{|\La_\varepsilon|^{k-i}}{(k-i)!}
\underset{\{x_1,\ldots,x_k\}\in
(\mathbb{R}^d)^k}{\textnormal{sup}}|F(x_1,\ldots,x_k)| \rightarrow 0
\end{align}
if\,$ a\rightarrow 0$ as \textnormal{(mess\,$ \Lambda_\eta (a)
\rightarrow 0)$} (see remark 5.2). Estimate \eqref{65} is a
consequence of \eqref{72} - \eqref{74}. The lemma is proven.
$$
\mspace{675mu}\blacksquare
$$

\end{document}